\documentstyle[12pt]{article}

\newtheorem{theorem}{Theorem}
\newtheorem{lemma}{Lemma}

\begin{document}

\newcommand{\lap}{\bigtriangleup}
\def\be{\begin{equation}}
\def\ee{\end{equation}}
\def\bea{\begin{eqnarray}}
\def\eea{\end{eqnarray}}
\def\beas{\begin{eqnarray*}}
\def\eeas{\end{eqnarray*}}
\def\n#1{\vert #1 \vert}
\def\nn#1{{\Vert #1 \Vert}}

\def\R{{\rm I\kern-.1567em R}}
\def\N{{\rm I\kern-.1567em N}}

\def\supp{\mbox{\rm supp}\,}
\def\suppi{\mbox{\scriptsize supp}\,}
\def\ekin{E_{\rm kin}}
\def\epot{E_{\rm pot}}

\def\D{{\cal D}}
\def\C{{\cal C}}
\def\X{{\cal X}}
\def\F{{\cal F}}
\def\P{{\cal P}}

\def\prf{\noindent
         {\em Proof :\ }}

\def\prfe{\hspace*{\fill} $\Box$

\smallskip \noindent}

\title{
Flat steady states in stellar dynamics---existence and stability}
\author{Gerhard Rein\\
        Mathematisches Institut\\
        der Universit\"at M\"unchen\\
        Theresienstr. 39\\
        80333 M\"unchen, Germany\\
        e-mail: rein@rz.mathematik.uni-muenchen.de}
\date{}
\maketitle
\begin{abstract}
We consider a special case of the three dimensional Vlasov-Poisson 
system where the particles are restricted to a plane, a situation 
that is used in astrophysics to model extremely flattened galaxies. 
We prove the existence of steady states of this system.
They are obtained as minimizers of an energy-Casimir functional from 
which fact a certain dynamical stability property is deduced. 
From a mathematics point of view these steady states provide 
examples of partially singular solutions of the three dimensional
Vlasov-Poisson system.

\end{abstract}

\section{Introduction}
\setcounter{equation}{0}

In astrophysics the time evolution of large stellar systems such as 
galaxies is often modeled by the Vlasov-Poisson system: 
\[
\partial_t f + v \cdot \nabla _x f - \nabla _x U \cdot 
\nabla _v f = 0,
\]
\[ 
\lap U = 4 \pi\, \rho,\ \lim_{x \to \infty} U(t,x)=0, 
\]
\[ 
\rho(t,x)= \int_{\R^3} f(t,x,v)dv .
\]
Here $f = f(t,x,v)\geq 0$ denotes the density of the stars
in phase space, $t \in \R$ denotes time, $x, v \in \R^3$ denote 
position and velocity respectively, $\rho$ is the spatial mass 
density, and $U$ the gravitational potential. The only interaction
between the stars is via the gravitational field which the
stars create collectively, in particular, collisions are neglected.
When modeling an extremely flattened galaxy the stars can be 
taken to be concentrated in a plane (the $(x_1,x_2)$-plane). 
The corresponding potential which is given by the usual integral 
representation induces a force field which accelerates the particles
only parallelly to the plane, and the Vlasov-Poisson system takes the form
\be \label{vlasov}
\partial_t f + v \cdot \nabla _x f - \nabla _x U \cdot 
\nabla _v f = 0,\ t \in \R,\ x, v \in \R^2,
\ee
\be \label{potential}
U(t,x)= - \int_{\R^2} \frac{\rho(t,y)}{\n{x-y}}dy,
\ee
\be \label{rho}
\rho(t,x)= \int_{\R^2} f(t,x,v)dv .
\ee
Note that from this point on,
$x, v \in \R^2$.
The three dimensional phase space and spatial densities are given
as
\[
\tilde f(t,x,x_3,v,v_3)=f(t,x,v)
\delta (x_3) \delta (v_3)
\]
and
\[
\tilde \rho (t,x,x_3)=\rho (t,x)
\delta (x_3) 
\]
where $\delta$ denotes the Dirac distribution.
It should be emphazised that the system (\ref{vlasov}), (\ref{potential}),
(\ref{rho}) is not a two dimensional version of the Vlasov-Poisson
system but a special case of the three dimensional system with 
partially singular phase space density.
In the present paper we are concerned with the existence of
steady states of this system and with their stability properties.
There are a number of aspects which make this problem interesting.
Although such flat solutions of the Vlasov-Poisson system 
occur as models in the astrophysics literature, cf.\
\cite{BT,FP}, we know of no mathematical investigation
of this situation. The fact that the distribution function is singular
in the $x_3$-direction, or, alternatively, that the two dimensional
Vlasov equation is coupled to a potential with the three dimensional
$1/\n{x}$-singularity, makes this problem mathematically nontrivial.
We refer to \cite{MZ}, where solutions of the Vlasov-Poisson
system which are measures
are treated in the one dimensional case; an extension of these
results to higher dimensions is not known. Finally, the method that
we employ to study the existence and the stability properties of
steady states was recently used in a spherically symmetric, regular,
three dimensional
situation in \cite{GR}. The present paper demonstrates that this
method extends beyond the case of spherical symmetry, although this assumption 
played an important role in \cite{GR}.

To see how steady states of the system 
(\ref{vlasov}), (\ref{potential}), (\ref{rho}) 
can be obtained,  note first that if $U=U(x)$ is time independent,
the particle energy
\be \label{parten}
E=\frac{1}{2}|v|^2 + U (x) 
\ee
is conserved along characteristics of (\ref{vlasov}).
Thus any function of the form
\be \label{ansatz}
f(x,v)=\phi(E)
\ee
satisfies the Vlasov equation.
We construct steady states as minimizers of
an appropriately defined energy-Casimir functional. Given a function
$Q=Q(f) \geq 0,\ f\geq 0$, we define
\[
\D(f) :=
\int\!\!\int Q(f)\,dv\,dx +{1\over 2} \int\!\!\int |v|^2 f\,dv\,dx
+\frac{1}{2} \int \rho_f U_f \,dx.
\]
Here $f=f(x,v)$ is taken from some appropriate set $\F_M$
of functions which in particular have total mass equal to a 
prescribed constant $M$, 
$\rho_f$ denotes the spatial density induced
by $f$ via (\ref{rho}), and $U_f$ denotes the potential induced by 
$\rho_f$ via (\ref{potential}).
If one can show that the functional $\D$ has a
minimizer, then the corresponding Euler-Lagrange equation turns out
to be of the form (\ref{ansatz}). 
An alternative way to obtain steady states would be to substitute 
(\ref{ansatz}) into (\ref{rho}) so that $\rho$ would become a functional
of $U$, and it would remain to solve (\ref{potential}) which becomes
a nonlinear integral equation for $U$. This route is followed
for example in \cite{BFH} for the regular three dimensional problem.
The major difficulty then is to show that the resulting
steady state has finite mass and compact support---properties which
are essential for a steady state to qualify as a physically viable model---, 
and this problem has been dealt with for the polytropic ansatz
\[
f(x,v)=(E-E_0)_+^\mu
\]
where $E_0$ is a constant, $-1 < \mu < 7/2$, and $(\cdot)_+$ denotes the positive part. Our approach has the advantage that
finiteness of the total mass and compact support are built in or
appear naturally, and these properties do not depend on a specific
ansatz like the polytropic one. Furthermore, the fact that the steady state
is obtained as a minimizer of the functional $\D$ implies a
certain nonlinear stability property of that steady state. 

The paper proceeds as follows. In the next section
the assumptions on the function $Q$ which determines our
energy-Casimir functional are stated, and some preliminary results, in
particular a lower bound of $\D$ on $\F_M$, are established. 
The main difficulties in finding a minimizer of $\D$
arise from the fact that $\D$ is neither positive definite 
nor convex, and from the lack of compactness: Along
a minimizing sequence some mass might escape to infinity. However, 
using the scaling properties of $\D$ and a certain splitting
estimate we show that
along a minimizing sequence the total mass has to concentrate in
a disc of a certain radius $R_M$, depending only on $M$.
In \cite{GR} the corresponding argument required the assumption of
spherical symmetry. In the present paper we only require axial symmetry
with respect to the $x_3$-axis.  
The corresponding estimates are proved in Section 3 
and are used in Section 4 to show 
the existence of a minimizer. It is then straight forward to
show that the Euler-Lagrange equation is (equivalent to) (\ref{ansatz}),
thereby completing the existence proof for the steady states.
The resulting stability property of such steady states is discussed in
Section 5. Since we have to restrict our functions in the set $\F_M$
to axially symmetric ones stability holds only with respect to such 
perturbations, and also perturbations transversal to the $(x_1,x_2)$-plane
are not covered. Moreover, the stability result is only conditional 
in the sense
that so far no existence theory for the initial value problem for
the flat Vlasov-Poisson system is available. To obtain a complete stability
result, global existence of solutions which preserve the energy-Casimir functional $\D$ would be needed, at least for data close to the steady
states. In the last section we briefly discuss the regularity
properties of the obtained steady states.

We conclude this introduction with some references to the 
literature. In the regular three dimensional situation
the existence of global classical solutions to the corresponding 
initial value problem has been 
shown in \cite{P}, cf.\ also \cite{H,LP,S}. The existence
of steady states for the case of the polytropic ansatz
was investigated in \cite{BFH} and \cite{BP}. We refer to \cite{FP}
for contributions to the stability problem in the astrophysics
literature. As to
mathematically rigorous results on the stability problem, we mention
\cite{G3,GR} for applications of the present approach in the
regular, three dimensional case,
cf.\ also \cite{Wo}.  An investigation
of linearized stability is given in \cite{BMR}. For the plasma physics case,
where the sign in the Poisson equation is
reversed, the stability problem is much easier and better
understood. We refer to \cite{BRV,GS1,GS2,R2}. A plasma physics situation
with magnetic field is investigated in \cite{G2}.

\section{Preliminaries; a lower bound for $\D$}
\setcounter{equation}{0}

We first state the assumptions on $Q$ which we need in the following:

\smallskip
\noindent
{\bf Assumptions on $Q$}: For $Q \in C^1([0,\infty[)$, $Q \geq 0$, and
constants $C_1, \ldots ,C_4 >0$, $F_0 >0$, and 
$0 < \mu_1,\,\mu_2,\,\mu_3 < 1$
consider the following assumptions:
\begin{itemize}
\item[(Q1)]
$Q(f) \geq C_1 f^{1+1/{\mu_1}},\ f \geq F_0$,
\item[(Q2)]
$Q(f) \leq C_2 f^{1+1/{\mu_2}},\ 0 \leq f \leq F_0$,
\item[(Q3)]
$Q(\lambda f) \geq \lambda^{1+1/{\mu_3}} Q(f),\ f \geq 0,\ 
0 \leq \lambda \leq 1$,
\item[(Q4)]
$Q''(f) > 0,\ f > 0$, and 
$Q'(0) = 0$.
\item[(Q5)]
$C_3  Q''(f) \leq Q''(\lambda f) 
\leq C_4 Q''(f)$ for $f > 0$ and $\lambda$
in some neighborhood of 1.
\end{itemize}
The above assumptions imply that $Q'$ is strictly
increasing with range $[0,\infty[$, and we denote its inverse
by $q$, i.~e., 
\be \label{qdef}
Q'(q(\epsilon)) = \epsilon,\ \epsilon\geq 0;
\ee
we extend $q$ by $q(\epsilon)=0,\ \epsilon < 0$.

\smallskip
\noindent
{\bf Remark:} The steady states obtained later will be of the
form 
\[
f_0 (x,v) = q(E_0 -E) 
\]
with some $E_0<0$ and $E$ as defined in 
(\ref{parten}).
If we take $Q(f) = f^{1+1/\mu},\ f\geq 0$, this leads to the
polytropic ansatz, and such a $Q$ satisfies
the assumptions above if $0<\mu<1$. If we take
\be \label{nopol}
Q(f) = C_1 f^{1+1/{\mu_1}} +  C_2 f^{1+1/{\mu_2}}
\ee
with $0 < \mu_1,\,\mu_2 < 1$ and constants
$C_1, C_2 >0$  then again the above assumptions hold, but $q$ is not
of polytropic form. Due to the assumption of axial symmetry
which we will have to make for other reasons,
\[
L_3=x_1v_2 - x_2 v_1,
\]
the $x_3$-component of angular momentum, is conserved along characteristics
as well.
It would be a purely technical matter to allow for depence on $L_3$ of a type
where for example the constants  in (\ref{nopol}) could be replaced by
functions of $L_3$ which are bounded and bounded away from 0. We refer to
\cite{GR} for the necessary modifications.

For a measurable function $f=f(x,v)$ we define
\[
\rho_f (x):= \int f(x,v)\, dv
\]
and 
\[
U_f := - \frac{1}{\n{\cdot}} \ast \rho_f;
\]
as to the existence of this convolution see Lemma~\ref{potest} below.
Then define
\beas
\ekin (f)
&:=&
\frac{1}{2} \int\!\!\int \n{v}^2 f(x,v)\,dv\,dx,\\
\epot (f)
&:=&
\frac{1}{2} \int \rho_f (x)\, U_f (x)\, dx =
- \frac{1}{2} \int\!\!\int \frac{\rho_f(x)\, \rho_f(y)}{\n{x-y}}dx\,dy,\\
\C(f)
&:=&
\int\!\!\int Q(f(x,v))\,dv\,dx,\\
\P(f)
&:=&
\ekin(f) + \C(f),\\
\D(f)
&:=&
\P(f) +\epot (f) .
\eeas
The sum $\ekin(f) + \epot(f)$ is the total energy corresponding to $f$, a 
conserved quantity for the time dependent problem,
and the same is true for $\C$, a Casimir functional
which corresponds to the conservation of phase space volume.
$\D$ is the energy-Casimir functional,
and $\P$ is the positive part of that functional.
We will also use the notation $U_\rho$ and $\epot(\rho)$ if
$\rho=\rho(x)$ is not necessarily induced by some $f=f(x,v)$.
The following two sets will serve as domains of definition
for the energy-Casimir functional $\D$: 
\be \label{spacedef}
\F_M := \left\{ f \in L^1(\R^4) 
\mid
f \geq 0,\ \int\!\!\int f dv\,dx = M,\
\P(f) < \infty\right\} ,
\ee
where $M>0$ is prescribed, and 
\be \label{sspacedef}
\F_M^S := \Bigl\{ f \in \F_M 
\mid f\ \mbox{is axially symmetric}\Bigr\}.
\ee
Here axial symmetry means that
\[
f(Ax,Av)=f(x,v),\ x,v \in \R^2,\ A \in \mbox{\rm SO}(2).
\]
When viewed as a function on the effective phase space
$\R^4$, $f$ is spherically symmetric, but when viewed as a
function over the full phase space $\R^6$, $f$ is only axially symmetric.
The induced potential does not share the properties of
spherically symmetric potentials which is why we prefer the
phrase axially symmetric. We do not restrict ourselves to the set
$\F_M^S$ from the beginning in order to point out where exactly the symmetry
is needed.

The aim of the present section is to establish a lower bound
for $\D$ of a form that will imply the boundedness of $\P$
along any minimizing sequence. 

\begin{lemma} \label{rhoest}
Let (Q1) hold and let $n_1 = 1 +\mu_1$. 
Then there exists a constant
$C>0$ such that for all $f\in \F_M$,
\[
\int \rho_f^{1+1/n_1} dx\le C \, ( M + \P(f)) . 
\]
\end{lemma}

\prf
We split the $v$ integral into small and large $v$'s and optimize
to obtain the estimate
\[
\rho_f (x) \leq
C \left(\int f^{1+1/\mu_1} dv \right)^{2 \mu_1/(4+2\mu)}
\left(\int |v|^2 f\, dv\right)^{2/(4+2\mu_1)}.
\]
By definition of $n_1$ and  assumption (Q1) we find
\beas
\int\rho_f^{1+1/n_1} dx
&\leq&
C \left(\int f^{1+1/\mu_1} dv\, dx +
\int |v|^2 f\, dv\, dx \right)\\
&\leq&
C\left(F_0^{1+1/\mu_1} \int f dv\, dx  + \frac{1}{C_1}
\int Q(f)\, dv\,dx + \int |v|^2 f\, dv\, dx \right),
\eeas
and by definition of $\P$ this is the assertion. \prfe

Note that $3/2 < 1+ 1/n_1 < 2$, and since by definition 
$\rho_f \in L^1 (\R^2)$ for $f \in \F_M$, we have $\rho_f \in L^{4/3}(\R^2)$
for $f\in \F_M$.

\begin{lemma} \label{potest}
If $\rho \in L^{4/3}(\R^2)$ then $U_\rho \in L^4(\R^2)$,
and there exists a constant $C>0$ such that for all $\rho \in L^{4/3}(\R^2)$,
\[
\nn{U_\rho}_4 \leq C \nn{\rho}_{4/3},\ -\epot(\rho)\leq C \nn{\rho}_{4/3}^2. 
\]
\end{lemma}

\prf
The assertion follows from
generalized Young's inequality
\cite[p.~32]{RS}, since $1/\n{\cdot} \in L^2_w(\R^2)$, the weak $L^2$-space,
and from H\"older's inequality. \prfe

Combining the previous two lemmata yields the desired lower bound
of $\D$ over the set $\F_M$:
 
\begin{lemma} \label{lower}
Let $Q$ satisfy assumption (Q1).
Then 
\[
\D_M := \inf\, \{\D(f) \mid f \in \F_M\} > - \infty,
\]
and there exists a constant $C_M >0$ depending on $M$ such that
\[
\D (f) \geq \P(f) - C_M \Bigl(1+\P(f)^{n_1/2}\Bigr),
\ f \in \F_M,
\]
and for any minimizing sequence $(f_n)\subset \F_M$ of $\D$ we have
\[
\P(f_n) \leq C_M,\ n \in \N.
\]
\end{lemma}

\prf
If we interpolate the $L^{4/3}$-norm
between the $L^1$-norm and the $L^{1+1/n_1}$-norm and apply 
Lemma~\ref{rhoest} we find
\[
\int \rho_f^{4/3} dx 
\leq
C_M \left(\int\rho_f^{1+1/n_1}dx\right)^{n_1/3} 
\leq
C_M \left(1+\P(f)\right)^{n_1/3} .
\]
Thus by Lemma~\ref{potest}
\[
\D(f) \geq \P(f) - C_M \Bigl(1+\P(f)\Bigr)^{2\frac{n_1}{3}\frac{3}{4}}
\geq \P(f) - C_M \Bigl(1+\P(f)^{n_1/2}\Bigr).
\]
Since $n_1<2$ the rest of the lemma is obvious
after possibly choosing $C_M$ larger. \prfe 

In later sections we will have to assume axial symmetry, and we will need the
fact that
\[
\D_M^S := \inf\, \{\D(f) \mid f \in \F_M^S\} > - \infty,
\]
which of course follows from the previous lemma.

\section{Scaling and Splitting}
\setcounter{equation}{0}

The behaviour of $\D$ and $M$ under scaling transformations can be
used to relate the $\D_M$'s for different values of $M$:

\begin{lemma} \label{scaling}
Let $Q$ satisfy the assumptions (Q1)--(Q3). Then 
$-\infty < \D_M  < 0$ for each $M>0$,
and for all $0< M_1 \leq M_2$,  
\[
\D_{M_1}\ge \left( {{M_1}\over{M_2}}\right )
^{1+\alpha}\D_{M_2}, 
\]
where $\alpha=1/(1-\mu_3)>0$.
The same assertions hold for $\D_M^S$ instead of $D_M$.
\end{lemma}

\prf
Given any function $f(x,v)$, we define a rescaled function 
 $\bar f(x,v)=af(bx,cv)$, where $a,\ b,\ c >0$. Then
\be \label{mscale}
\int\!\!\int \bar f\,dv\,dx = ab^{-2}c^{-2}\int\!\!\int f\,dv\,dx
\ee
and
\be
\D(\bar f) =
b^{-2}c^{-2} \C(af) + a b^{-2} c^{-4} \ekin(f)
+ a^2b^{-3} c^{-4}\epot (f) . \label{dscale}
\ee

\noindent
{\em Proof of $\D_M <0$:} 
Fix some $f\in \F_1^S$ with compact support and $f\leq F_0$,
and let $a=M b^2c^2$
so that $\bar f \in \F_M^S$.
The last term in $\D(\bar f)$ is negative and of the order $b$,
and we want to make this term dominate the others as $b \to 0$.
Choose  $c=b^{-\gamma/2}$ so that $a=M b^{2-\gamma}$,
and assume that $a \leq 1$ so that $af \leq F_0$. By (Q2),
\[
\D(\bar f) \leq
C \left( b^{(2-\gamma)/\mu_2} + b^\gamma \right) -\overline{C} b 
\]
where $C,\ \overline{C} >0$ depend on $f$. Since
we want the last term to dominate
as $b \to 0$, we need $\gamma >1$ and $(2-\gamma)/\mu_2 >1$,
and, in order that $a \leq 1$ as $b \to 0$, also
$\gamma < 2$. Such a choice of $\gamma$ is possible since $\mu_2 < 1$,
and thus $\D (\bar f) < 0$ for $b$ sufficiently small. 

\noindent
{\em Proof of the scaling inequality:}
Assume that $f \in \F_{M_2}$ and $\bar f \in \F_{M_1}$ so that by 
(\ref{mscale}),
\be \label{m1m2}
a b^{-2}c^{-2} = \frac{M_1}{M_2} =: m \leq 1.
\ee
By (\ref{dscale}) and (Q3),
\[
\D(\bar f) \geq m a^{1/\mu_3} \C(f) + m c^{-2} \ekin (f)
+ m^2 b \epot(f)
\]
provided $a \leq 1$. Now we require that
\[
m a^{1/\mu_3} = m c^{-2} = m^2 b.
\]
Together with (\ref{m1m2}) this determines $a,\ b,\ c$ in terms of $m$.
In particular,
\[
a=m^{\mu_3/(1-\mu_3)} \leq 1
\]
as required and
\[
\D(\bar f) \geq m^{1+\frac{1}{1-\mu_3}}\D(f).
\]
Since for any given choice of $a,\ b,\ c$ the mapping $f \mapsto \bar f$
is one-to-one and onto between $\F_{M_2}$ and $\F_{M_1}$
as well as between  $\F_{M_2}^S$ and $\F_{M_1}^S$
the scaling inequality follows. \prfe

The following two lemmata are crucial in proving that along
a minimizing sequence the mass concentrates in a certain ball.
It is here that we need the additional symmetry assumption
and where the estimates become more involved than in the regular
spherically symmetric case. The aim is to estimate the effect on $\D$
of splitting the matter distribution into a part inside
a ball $B_R$ of (large) radius $R$ about $0$ and a part outside. 
  
\begin{lemma} \label{interpot}
There exists a constant $C>0$ such that for every
$\rho \in L^1\cap L^{4/3}(\R^2)$ which is nonnegative and axially symmetric,
i.~e., $\rho (x) = \rho (\n{x})$, and every $R>0$ the following estimate holds:
\[
- \int_{\n{x}>R} \rho(x)\,U_\rho(x)\,dx \leq C R^{-1/2} \nn{\rho}_{4/3} 
\int_{\n{x}>R} \rho(x)\, dx .
\]
\end{lemma}

\prf
Due to the symmetry of $\rho$
the potential is given by
\[
U_\rho (x)=U_\rho (r)=- 4 \int_0^\infty \frac{s}{r+s} \rho(s)\, 
K\left(\frac{2 \sqrt{rs}}{r+s}\right)\,ds 
\]
where the elliptic integral $K$ is defined as
\[
K(\xi) = \int_0^{\pi/2} \frac{d\phi}{\sqrt{1-\xi^2 \sin^2\phi}}
= \int_0^1 \frac{dt}{\sqrt{1-\xi^2 t^2}
\sqrt{1-t^2}},\ 0 \leq \xi <1.
\]
We need to estimate the singularity in $K$:
\beas
K(\xi) 
&\leq&
\int_0^1  \frac{dt}{\sqrt{1-\xi t} \sqrt{1-t}}
= \frac{1}{\sqrt{\xi}} \ln\frac{1+\sqrt{\xi}}{1-\sqrt{\xi}}\\ 
&\leq&
C\left(1-\ln (1-\xi)\right),\ 0\leq \xi <1.
\eeas
Substituting for $\xi$ yields
\[
1-\xi = \frac{(\sqrt{r} - \sqrt{s})^2}{r+s}
= \frac{s}{r+s} \left(1-\sqrt{r/s}\right)^2 
\geq \frac{1}{2} \left(1-\sqrt{r/s}\right)^2\geq \frac{1}{8} (1-r/s)^2
\]
for $0 \leq r \leq s$; the case $r\geq s$ is analogous. Thus
\be \label{kest}
K\left(\frac{2 \sqrt{rs}}{r+s}\right)
\leq
C \left( 1 - \ln (1-[r,s])\right),\ r,s>0
\ee
where
\[
[r,s]:= \min\left\{ \frac{r}{s},\frac{s}{r}\right\}.
\]
Now
\[
- \int_{\n{x}>R} \rho(x)\,U_\rho(x)\,dx
= 8 \pi \int_R^\infty\int_0^\infty \rho(r)\,\rho(s)\,\frac{rs}{r+s} 
K\left(\frac{2 \sqrt{rs}}{r+s}\right)\,dr\,ds
=I_1 + I_2
\]
where in $I_1$ the variable $r$ ranges in $[0,2s]$ and in $I_2$ it ranges in
$[2s,\infty[$. Using (\ref{kest}) and H\"older's inequality we find
\beas
I_1 
&\leq&
C \int_R^\infty \rho(s) \int_0^{2s} r \rho(r) 
\left(1-\ln (1-[r,s])\right)\,dr\,ds\\
&\leq&
C \nn{\rho}_{4/3} \int_R^\infty\rho(s)\left(\int_0^{2s} r 
\Bigl(1-\ln (1-[r,s]) \Bigr)^4 dr\right)^{1/4}ds\\
&\leq&
C \nn{\rho}_{4/3} \int_R^\infty s^{1/2} \rho(s)\,ds
\leq
C \nn{\rho}_{4/3} R^{-1/2} \int_R^\infty s \rho(s)\,ds;
\eeas
note that with $\sigma = r/s$,
\beas
\int_0^{2s} r 
\Bigl(1-\ln (1-[r,s])\Bigr)^4 dr
&=&
s^2 \int_0^1 \sigma
\Bigl(1-\ln (1-\sigma)\Bigr)^4 d\sigma\\
&&
{} + s^2 \int_1^2 \sigma
\Bigl(1-\ln(1-1/\sigma)\Bigr)^4 d\sigma\\
&=& C s^2,\ s>0.
\eeas
The second term is much easier to estimate:
For $r>2s$ we have $- \ln (1-s/r) \leq \ln 2$,
and by H\"older's inequality,
\beas
I_2
&\leq&
C \int_R^\infty s \rho(s) \int_{2s}^\infty \frac{r}{r+s} \rho(r)\,dr\,ds
\leq
C \int_R^\infty  s \rho(s)\,ds \int_R^\infty \rho(r)\,dr\\
&\leq&
C \nn{\rho}_{4/3} R^{-1/2} \int_R^\infty s\rho(s)\,ds .
\eeas
Together with the estimate for $I_1$ this completes the proof. \prfe

\begin{lemma} \label{split}
Let $Q$ satisfy the assumptions (Q1)--(Q3) and let $f\in \F_M^S$.
Then
\[
\D (f)- \D^S_M
\geq
\left(\frac{C_\alpha  \D_M^S}{M^2} \int_{\n{x}<R}\int f\,dv\,dx
 - \frac{C_M}{\sqrt{R}} \right)
\int_{\n{x}>R}\int f\,dv\,dx ,
\ R>0,
\]
where the constant $C_\alpha < 0$ depends on $\alpha$ from Lemma~\ref{scaling}
and $C_M>0$ depends on $M$.
\end{lemma}

\prf 
Let $B_R$ denote the ball of radius $R$ about $0$ in $\R^2$, let 
$1_{B_R\times \R^2}$ be the characteristic function of $B_R \times \R^2$,
\[
f_1=1_{B_R\times \R^2} f,\ f_2 = f - f_1,
\]
and let $\rho_i$ and $U_i$ denote the induced spatial densities
and potentials respectively, $i=1,2$. We abbreviate $\lambda=\int f_2$.
Then
\beas
\D (f)
&=&
\P(f_1)+\P(f_2)
+ \frac{1}{2} \int U_1\,\rho_1\,dx +  \frac{1}{2} \int U_2\,\rho_2\,dx
+ \int U_1\,\rho_2\,dx\\
&\geq&
\D^S_{M-\lambda} + \D^S_{\lambda}
- C_M R^{-1/2} \lambda
\eeas
since $f_1 \in \F_{M-\lambda}^S$ and $f_2 \in \F_\lambda^S$. 
To estimate the ``mixed term'' in the potential energy
we have used Lemma~\ref{interpot}; note that for $f \in \F_M$,
$\nn{\rho_f}_{4/3}$ is bounded by a constant depending
only on $M$, cf.\ Lemma~\ref{rhoest}.
Since $\alpha>0$, there is a constant $C_\alpha <0$, such that 
\[
(1-x)^{1+\alpha}+x^{1+\alpha}-1\le C_\alpha (1-x)x,\ 0\le x\le 1. 
\]
Using Lemma~\ref{scaling}  and
noticing that $\D^S_M<0$ we find that
\beas
\D (f) - \D^S_M
&\geq&
\left[(1-\lambda/M)^{1+\alpha}
+(\lambda/M)^{1+\alpha}-1 \right] \D^S_M - C_M R^{-1/2} \lambda\\
&\geq&
C_\alpha \D^S_M 
\left(1-\lambda/M\right)\lambda/M  
- C_M R^{-1/2} \lambda
\eeas
which is the assertion. \prfe

\section{Minimizers of $\D$}
\setcounter{equation}{0}

Before we show the existence of a minimizer of $\D$ over the set $\F_M^S$
we use Lemma~\ref{split} to show that along
a minimizing sequence the mass has to concentrate in a certain ball:

\begin{lemma} \label{r0}
Let $Q$ satisfy the assumptions (Q1)--(Q3), and define
\[
R_M:= \left(\frac{2 M C_M}{C_\alpha \D^S_M}\right)^2
\]
where $C_\alpha <0$ and $C_M>0$ are as in Lemma~\ref{split}.
If $(f_n) \subset \F_M^S$ is a minimizing sequence of $\D$, then
for any $R > R_M$,
\[
\lim_{n\to \infty} \int_{|x|\ge R} \int f_n dv\,dx = 0.
\]
\end{lemma}

\prf 
If not, there exist some $R>R_M$, 
$\lambda>0$, and a subsequence, called $(f_n)$ again, such that 
\[
\lim_{n\to \infty}
\int_{|x|\ge R}\int f_n dv\,dx = \lambda .
\]
For every $n\in \N$ we can now choose $R_n>R$ such that
\[
\lambda_n := \int_{|x|\geq R_n}\int f_n dv\,dx = {1\over 2}
\int_{|x|\ge R}\int f_n dv\,dx .
\]
Then 
\[
\lim_{n\to \infty}
\int_{|x|\ge R_n}\int f_n dv\,dx =
\lim_{n\to\infty}
\lambda_n = \lambda/2>0.
\]
Applying Lemma~\ref{split} to $B_{R_n}$ we get 
\begin{eqnarray*}
\D (f_n) - \D^S_M
&\geq& 
\left(\frac{C_\alpha  \D_M^S}{M^2}  (M-\lambda_n) -
\frac{C_M}{\sqrt{R_n}}\right) \lambda_n
> \left({{C_\alpha  \D_M^S}\over
{M^2}}  (M-\lambda_n) -\frac{C_M}{\sqrt{R}}\right) \lambda_n\\
&\to&
\left(\frac{C_\alpha  \D_M^S}{M^2}  (M-\lambda/2) -
\frac{C_M}{\sqrt{R}}\right) \frac{\lambda}{2}
\geq 
\left(\frac{C_\alpha  \D_M^S}{2 M}  -
\frac{C_M}{\sqrt{R}}\right) \frac{\lambda}{2} 
\end{eqnarray*}
as $n\to\infty$, since $0 < \lambda \leq M$. 
By definition of $R_M$ the expression in the
parenthesis is positive for $R>R_M$, and
this contradicts the fact that $(f_n)$ is a minimizing 
sequence. \prfe

As a further prerequisite for the existence proof of a minimizer
we establish a compactness property of the potential energy functional:

\begin{lemma} \label{compact}
Let $(\rho_n)\subset L^{3/2} \cap L^1 (\R^2)$ be bounded and axially symmetric
with
\[
\rho_n \rightharpoonup \rho_0 \ \mbox{weakly in}\ L^{3/2}(\R^2),
\ n \to \infty.
\]
Then
\[
\epot(\rho_n -\rho_0) \to 0 \ \mbox{and}\ 
\epot(\rho_n) \to \epot(\rho_0),\ n \to \infty.
\]
\end{lemma}

\prf
We consider the convergence of $\epot(\rho_n -\rho_0)$ first.
By Lemma~\ref{interpot}
and H\"older's inequality 
\beas
\n{\epot(\rho_n -\rho_0)}
&\leq&
\left|\int (U_{\rho_n}-U_{\rho_0}) (\rho_n - \rho_0) \,dx\right|\\
&\leq&
\nn{\rho_n-\rho_0}_{4/3} \nn{U_{\rho_n,R}-U_{\rho_0,R}}_{L^4(B_R)} + \frac{C}{\sqrt{R}},
\eeas
for any $R>0$, where 
\[
U_{\rho,R} (x):= -\int_{\n{y}\leq R} \frac{\rho(y)}{\n{x-y}}dy,\ x \in \R^2.
\]
Thus it suffices to show that for $R>0$ fixed the mapping
\[
T : L^{3/2} (B_R) \ni \rho \mapsto 1_{B_R} (\rho \ast k) \in
L^4(\R^2)
\]
is compact where $k:= 1_{B_{2R}} 1/\n{\cdot}$; 
note that we may cut off the Green's function
as indicated since only $x,y$ with $\n{x},\n{y} \leq R$ need to
be considered.
We use the Frech\'{e}t-Kolmogorov criterion to show that $T$ is compact.
Let $S\subset  L^{3/2} (B_R)$ be bounded. 
Then $TS$ is bounded in $L^4(\R^2)$ by Lemma~\ref{potest}. Since the
elements in $TS$ have a uniformly compact support it remains to
show that for $\rho \in S$,
\[
(T\rho)_h \to T\rho \ \mbox{in}\ L^4 (\R^2),\ h \to 0,
\]
where $g_h:=g(\cdot +h),\ h\in \R^2$. But by
Young's inequality,
\[
\nn{(T\rho)_h - T\rho}_4 \leq \nn{\rho\ast(k_h-k)}_4
\leq \nn{\rho}_{3/2} \nn{k_h-k}_{12/7} \to 0
\]
uniformly on $S$ as $h \to 0$, since $k \in L^{12/7}(\R^2)$. Since
\[
\epot(\rho_n) - \epot(\rho_0)
= - \int U_{\rho_0} (\rho_n - \rho_0) - \epot(\rho_n-\rho_0)
\]
the fact that $U_{\rho_0} \in L^4(\R^2)$ together with the weak convergence of
$\rho_n$ implies the remaining assertion.
\prfe

\begin{theorem} \label{exminim}
Let $Q$ satisfy the assumptions (Q1)--(Q4), and
let $(f_n) \subset \F_M^S$ be a minimizing sequence of 
$\D$. Then there is a minimizer $f_0\in \F_M^S$ and a subsequence
$(f_{n_k})$ such that 
$\D (f_0) = \D_M^S$, $\supp f_0 \subset B_{R_M}\times \R^2$ with 
$R_M$ as in Lemma~\ref{r0},
and $f_{n_k} \rightharpoonup f_0$ weakly in 
$L^{1+1/\mu_1} (\R^4)$.
Furthermore,
$\epot(f_{n_k} -f_0) \to 0$.
\end{theorem}

\prf
By Lemma~\ref{lower}, $(\P(f_n))$ is bounded. Let
$p_1=1+1/\mu_1$. Then the sequence
$(f_n)$ is bounded in $L^{p_1} (\R^4)$ by assumption (Q1). 
Thus there exists a weakly convergent
subsequence, denoted by $(f_n)$ again, i.~e.,
\[
f_n \rightharpoonup f_0\ \mbox{weakly in }\ L^{p_1} (\R^4).
\]
Clearly, $f_0 \geq 0$ a.~e.,\ and $f_0$ is axially symmetric.
Since by Lemma~\ref{r0}
\beas
M 
&=& 
\lim_{n \to \infty} \int_{\n{x}\leq R_1}\int_{\n{v}\leq R_2}f_n dv\,dx
+ \lim_{n \to \infty} \int_{\n{x}\leq R_1}\int_{\n{v}\geq R_2}f_n dv\,dx\\
&\leq&
\lim_{n \to \infty} \int_{\n{x}\leq R_1}\int_{\n{v}\leq R_2}f_n dv\,dx
+ \frac{C}{R_2^2}
\eeas
where $R_1 > R_M$ and $R_2 >0$ are arbitrary, it follows that
\[
\int_{\n{x}\leq R_1}\int f_0 dv\,dx = M
\]
for every $R_1 > R_M$. This proves the assertion on $\supp f_0$
and $\int\!\!\int f_0 =M$. Also by weak convergence 
\be \label{wke}
\int\!\!\int \n{v}^2 f_0 dv\,dx \leq \liminf_{n \to \infty}
\int\!\!\int \n{v}^2 f_n dv\,dx < \infty.
\ee
By Lemma~\ref{rhoest} $(\rho_n)=(\rho_{f_n})$ is bounded in 
$L^{1+1/n_1} (\R^2)$ where
$n_1 = \mu_1 + 1$. After extracting a further subsequence, we thus have
that
\[
\rho_n \rightharpoonup \rho_0:=\rho_{f_0} \
\mbox{weakly in }\ L^{3/2} (\R^2),
\]
and Lemma~\ref{compact} implies the convergence of the potential energy
term.
 
It remains to show that $f_0$ is actually a minimizer, in
particular, $\P(f_0) < \infty$ so that $f_0 \in \F_M^S$.
By Mazur's Lemma there exists a sequence $(g_n) \subset L^{p_1}(\R^4)$ 
such that $g_n \to f_0$ strongly in $L^{p_1}(\R^4)$ and
$g_n$ is a convex combination of $\{f_k\mid k\geq n\}$.
In particular, $g_n \to f_0$ a.~e.\ on $\R^4$. By (Q4) the functional
$\C$ is convex. 
Combining this with Fatou's Lemma implies that
\[
\C(f_0) \leq 
\liminf_{n \to \infty} \C(g_n)
\leq \limsup_{n\to \infty} \C (f_n).
\]
Together with (\ref{wke}) this implies that
\[
\P(f_0) \leq \lim_{n\to \infty} \P(f_n) < \infty ;
\]
note that $\lim_{n\to \infty} \P(f_n)$ exists.
Therefore,
\[
\D(f_0) = \P(f_0) +\epot(f_0) 
\leq 
\lim_{n\to \infty} \left(\P(f_n) +\epot(f_n)\right)
= \D_M^S,
\]
and the proof is complete. \prfe

\begin{theorem} \label{propminim}
Let $Q$ satisfy the assumptions (Q1)--(Q5), and
let $f_0 \in \F_M$ be a minimizer of $\D$. Then
\[
f_0 (x,v) = 
q(E_0-E) \ \mbox{a.~e.\ on}\ \R^4,
\]
where
\[
E = \frac{1}{2} \n{v}^2 + U_0 (x),
\]
\[
E_0 = {1\over M} \int\!\!\int \left(Q'(f_0)
+ E \right)\,f_0\,dv\,dx  <0,
\]
$U_0$ is the potential induced by $f_0$, and $q$ is as defined in (\ref{qdef}).
\end{theorem}
Note that $U_0=U_{f_0}$ by construction, and $f_0$ is a function of
the particle energy only and thus a steady state of the system
(\ref{vlasov}),\ (\ref{potential}),\ (\ref{rho}). The regularity
of $U_0$ and thus the sense in which $f_0$ satisfies the
Vlasov equation (\ref{vlasov}) is investigated in the last section.

\prf 
Let $f_0$ be a minimizer.  
For fixed $\epsilon>0$ let $\eta : \R^4 \to \R$ be measurable,
with compact support, axially symmetric, and such that
\[
\n{\eta} \leq 1,\ \mbox{a.~e.\ on}\ \R^4,\   
\eta \geq 0\ \mbox{a.~e.\ on}\ \R^4 \setminus \supp f_0,
\]
and 
\[
\epsilon \leq f_0 \leq \frac{1}{\epsilon}\ 
\mbox{a.~e.\ on}\ \supp f_0 \cap \supp \eta .
\]
Below we will occasionally argue pointwise on $\R^4$ so we choose
a representative of $f_0$ satisfying the previous estimate pointwise
on $\supp f_0 \cap \supp \eta$. 
For 
\[
0 \le h \le {{\epsilon}\over{2(1+\|\eta\|_1)}}
\] 
we define 
\[
g(h)= M {{h\eta+f_0}\over {\|h\eta+f_0\|_1}}. 
\]
This defines a variation of $f_0$ with $g(h) \in \F_M^S$ and $g(0)=f_0$;
note that  
\[
M-\frac{\epsilon}{2} \leq \|h\eta+f_0\|_1
\leq M + \frac{\epsilon}{2}.
\]
We expand $\D(g(h))-\D(f_0)$ in powers of $h$:
\bea
\D(g(h))-\D(f_0)
&=&
\int\!\!\int \Bigl( Q(g(h))-Q(f_0)\Bigr)\,dv\,dx + 
{1\over 2} \int\!\!\int|v|^2(g(h)-f_0)\,dv\,dx \nonumber\\
&& 
{} + \int\!\!\int U_0 (g(h)-f_0)\,dv\,dx + \epot(g(h)-f_0) . 
\label{dexpan}
\eea
Since
$g(h)\ge 0$ on $\R^4$, $g(h)$ is differentiable
with respect to $h$, and we write $g'(h)$ for this
derivative. Note that both $g(h)$ and $g'(h)$ are functions
of $(x,v)\in \R^4$, but we suppress this dependence and obtain
\beas
g'(h)
&=& 
\frac{M}{\|h\eta+f_0\|_1}\eta - M 
\frac{h \eta + f_0}{\|h\eta + f_0\|_1^2} \int\!\!\int\eta\,dv\,dx, \\
g''(h)
&=&
- 2\frac{M}{\|h\eta + f_0\|_1^2} \left(\int\!\!\int \eta\,dv\,dx\right)\, \eta
+2 M \frac{h\eta + f_0}{\|h\eta + f_0\|_1^3} 
\left(\int\!\!\int\eta\,dv\,dx\right)^2 .
\eeas
Now
\be \label{gprime}
g'(0) = \eta - \frac{1}{M} \left(\int\!\!\int \eta\,dv\,dx\right)\, f_0
\ee
and
\[
\n{g''(h)} \leq C\,(\n{\eta} + f_0)
\]
so that on $\R^4$,
\[
\left|g(h) - f_0 - h g'(0)\right|
\leq C\,h^2 (\n{\eta} + f_0);
\]
in the following, constants denoted by $C$ may depend on $f_0$, $\eta$,
and $\epsilon$ but never on $h$.
We can now estimate the last three terms in (\ref{dexpan}):
\bea
\int\!\!\int|v|^2(g(h)-f_0)\,dv\,dx
&=&
h \int\!\!\int \n{v}^2 g'(0)\,dv\,dx + O(h^2), \label{term3}\\
\int\!\!\int U_0 (g(h)-f_0)\,dv\,dx 
&=&
h \int\!\!\int U_0\, g'(0)\,dv\,dx  + O(h^2),\label{term4}\\  
|\epot(g(h)-f_0)| 
&\leq&
C \nn{\rho_{g(h)}-\rho_0}_{4/3}^2 \leq C h^2. \label{term5}
\eea
For the last estimate we used Lemma~\ref{potest} and the fact that
\[
\n{\rho_{g(h)} (x) - \rho_0(x)} \leq C h \int (\n{\eta} + f_0)(x,v)\,dv.
\]
It remains to estimate the first term in (\ref{dexpan}). 
Consider first a point $(x,v) \in \supp f_0$ with $f_0(x,v)>0$. Then
\beas
Q(g(h))-Q(f_0)
&=&
Q'(f_0) (g(h)-f_0) + \frac{1}{2} Q''(\tau)
(g(h)-f_0)^2 \\
&=&
h Q'(f_0) g'(0) + 
h^2 \frac{1}{2} Q'(f_0) g''(\theta)
+ \frac{1}{2} Q''(\tau) (g(h)-f_0)^2
\eeas
where $\tau$ lies between $g(h)$ and $f_0$ and $\theta$ lies between
$0$ and $h$; both $\tau$ and $\theta$ depend on $(x,v)$.
Thus
\[
\Bigl| Q(g(h))-Q(f_0) - h Q'(f_0) g'(0)\Bigr| 
\leq
C Q'(f_0)\, (\n{\eta}+f_0)\, h^2
+ C Q''(\tau)  (\n{\eta}^2 +f_0^2)\, h^2.
\]
On $\supp f_0$ we have 
\[
\frac{1}{4}\,f_0 \le g(h) \le 2 f_0
\]
provided $0<\epsilon < \epsilon_0$ with $\epsilon_0>0$ sufficiently small.
Thus $\tau$ lies between $f_0/4$ and $2f_0$, and
by iterating (Q5) a finite, $h$-independent number of times 
we find
\[
Q''(\tau) \leq C  Q''(f_0).
\]
By (Q3) and (Q5),
\[
\left(2^{1+1/\mu_3} -1\right) Q(f_0)\geq
Q(2 f_0) - Q(f_0) \geq 
Q'(f_0)\, f_0 + C Q''(f_0)\, f_0^2
\]
and thus
\[
\Bigl| Q(g(h))-Q(f_0) - h Q'(f_0) g'(0)\Bigr|
\leq C Q(f_0) h^2 + C \n{\eta} h^2;
\]
here we used the continuity of $Q'$ and $Q''$ and
the fact that $\epsilon \leq f_0 \leq 1/\epsilon$ on  
$\supp \eta \cap \supp f_0$. 
The above estimate
holds for any point $(x,v) \in \supp f_0$ with $f_0(x,v)>0$.
Now consider a point $(x,v)$ with $f_0(x,v)=0$. Then 
\[
g(h)= M \frac{h\eta}{\|h\eta+f_0\|_1} \leq C \n{\eta} h
\]
so that by (Q4) and (Q2),
\beas
\Bigl| Q(g(h))-Q(f_0) - h Q'(f_0) g'(0)\Bigr|
= Q(g(h))
&\leq& 
Q(C h\n{\eta})\\
&\leq&
C \n{\eta}^{1+1/\mu_2} h^{1+1/\mu_2}
\eeas
for $h>0$ sufficiently small. Thus
\be \label{term1}
\int\!\!\int
\Bigl| Q(g(h))-Q(f_0) - h Q'(f_0) g'(0)\Bigr|\,dv\,dx
\leq C h^{1+\delta}
\ee
for some $\delta >0$.  
Combining (\ref{term3}), (\ref{term4}), (\ref{term5}),
and (\ref{term1}) with the fact that $f_0$ is a minimizer we find
\[
0 \leq \D(g(h)) - \D(f_0) =
h \int\!\!\int  
\left(Q'(f_0) + \frac{1}{2} \n{v}^2 + U_0 \right)
\,g'(0) \, dv\,dx  + O(h^{1+\delta})
\]
for all $h>0$ sufficiently small. Recalling
(\ref{gprime}) and the definitions of $E$ and $E_0$ this implies that
\[
\int\!\!\int \Bigl(Q'(f_0)
+ E - E_0 \Bigr)\, \eta\, dv\, dx \geq 0 .
\]
Recalling the class of admissable test functions $\eta$ 
and the fact that $\epsilon >0$ is arbitrary, provided it
is sufficiently small, we conclude that 
\[
E - E_0 \geq 0 \ \ \mbox{a.~e.\ on}\ \R^4 \setminus \supp f_0
\]
and
\[
Q'(f_0) +  E - E_0 = 0 \ \ \mbox{a.~e.\ on}\
\supp f_0.
\]
By definition of $q$---cf.\ (\ref{qdef})---this implies that
\[
f_0 (x,v) = q(E_0-E)\ \mbox{a.~e.\ on}\ \R^4.
\]
Since $\rho_0$ has compact support and 
$\lim_{x \to \infty} U_0 (x)=0$ we conclude that $E_0<0$.
\prfe

\section{Dynamical Stability}
\setcounter{equation}{0}

We now discuss the dynamical stability of $f_0$. 
As noted in the introduction the existence of solutions
to the initial value problem for the system (\ref{vlasov}), (\ref{potential}),
(\ref{rho}) is open. In the following we therefore have to assume that
for initial data in some (reasonably large) set
$\X \subset \F_M^S$ the system has a solution $f(t)$ with
$f(t) \in \F_M^S$ and $\D(f(t)) = \D(f(0))$, $t\geq 0$;
classical solutions of the regular three dimensional Vlasov-Poisson system 
have these properties.
The considerations below are only formal, 
and we emphasize this fact 
by not stating any theorems but only giving the stability
estimates. 
First we note
that for $f \in \F_M$,
\begin{equation}
\D (f)- \D (f_0)=d(f,f_0) +\epot(f-f_0)\label{d-d}
\end{equation}
where
\[
d(f,f_0) = \int\!\!\int \Bigl[Q(f)-Q(f_0)+(E-E_0)(f-f_0)\Bigr]\,dv\,dx.
\]
Next we observe that $d(f,f_0) \geq 0,\ f \in \F_M$.
For $E-E_0 \geq 0$ we have $f_0 = 0$, and thus
\[
Q(f)-Q(f_0)+(E-E_0)(f-f_0) \geq Q(f) \geq 0.
\]
For $E-E_0 <  0$,
\be \label{dexpansion}
Q(f)-Q(f_0)+(E-E_0)(f-f_0) = 
\frac{1}{2} Q''(\tilde f) (f - f_0)^2 \geq 0
\ee
provided $f>0$; here $\tilde f$ is between $f$ and $f_0$. If $f=0$,
the left hand side is still nonnegative by continuity. 
Now let $Q$ satisfy the assumptions (Q1)--(Q5) and
assume that the minimizer $f_0$ is unique in $\F_M^S$. Then we
obtain the following stability estimate:\\
{\em For every $\epsilon>0$ there is $\delta>0$ such that
for any solution $f(t)$ of the flat Vlasov-Poisson system
with $f(0) \in \X$,
\[
d(f(0),f_0) + \n{\epot(f(0)-f_0)} < \delta
\]
implies
\[
d(f(t),f_0) + \n{\epot(f(t)-f_0)} < \epsilon,
\ t \geq 0.
\] 
}
Assume this assertion were false. 
Then there exist $\epsilon_0>0$, $t_n>0$, and
$f_n(0) \in \X$ such that 
\[
d(f_n(0),f_0) + \n{\epot(f_n (0)-f_0)}
 = \frac{1}{n}
\]
but 
\[
d(f_n(t_n),f_0) +
\n{\epot(f_n(t_n)-f_0)} \ge \epsilon_0>0.
\]
From (\ref{d-d}), we have 
$\lim_{n\to\infty} \D (f_n(0))=\D_M^S$. 
Since  $\D (f)$ is invariant under the assumed Vlasov-Poisson flow,
\[
\lim_{n\to\infty} 
\D (f_n(t_n))=\lim_{n\to\infty} 
\D (f_n(0))=\D_M^S.
\]
Thus, $(f_n(t_n)) \subset \F_M^S$
is a minimizing sequence of $\D$, and by Theorem~\ref{exminim} ,
we deduce that---up to a 
subsequence---$\epot(f_n(t_n)-f_0)\to 0$. 
Again by (\ref{d-d}), 
$d(f_n(t_n),f_0)\to 0$, a contradiction.

Provided the assumed global Vlasov-Poisson flow is such that in addition
$\nn{f(t)}_\infty = \nn{f(0)}_\infty$, $t\geq0$, and that $Q$ is such that
\[
C_1 := \inf\left\{Q''(f) 
\mid 0 < f \leq C_0\right\}
>0
\]
for some constant $C_0 > \nn{f_0}_\infty$, then for
$f(0)\leq C_0$ one obtains the stability estimate 
\[
\int\!\!\int_{\R^4\setminus\suppi f_0} Q(f(t))\,dv\,dx
+ \frac{C_1}{2} \int\!\!\int_{\suppi f_0} \n{f(t)-f_0}^2 dv\,dx 
+ \n{\epot(f(t)-f_0)}< 
\epsilon.
\]
This follows by estimating $Q''$ in the expansion (\ref{dexpansion})
from below. 
 
If the minimizer $f_0$ of $\D$ is not unique (and not isolated)
in $\F_M^S$, then a solution starting close to $f_0$---in the sense
of the above measurement for the deviation---remains close to the set
of all minimizers in $\F_M^S$. In the regular, three dimensional case
uniqueness of the minimizer can be shown for the polytropic ansatz,
cf.\ \cite{GR}.

\section{Regularity}
\setcounter{equation}{0}

So far the steady states obtained in Section 4 satisfy the Vlasov-Poisson
system (\ref{vlasov}), (\ref{potential}), (\ref{rho}) in a 
rather weak sense, in particular, the potential need not be sufficiently
regular for characteristics of the Vlasov equation to exist so that the
precise meaning of $f_0$ being a function of an invariant
of the particle trajectories is questionable. The present section
will remedy this under some very mild additional assumptions:

\begin{theorem}
Assume that $Q$ satisfies conditions (Q1)--(Q5), and in addition
\[
Q'(f) \geq C_1 f^{1/\mu_1},\ f\geq F_0.
\]
Let $f_0\in \F_M^S$ be a minimizer of $\D$ as obtained in Theorem~1,
and $\rho_0,\ U_0$ the induced spatial density and potential respectively.
Then $\rho_0,\ U_0 \in C^1(\R^2)$, and the first derivatives of
$U_0$ are H\"older continuous. If the function $q$ defined in (\ref{qdef})
is locally H\"older continuous, then  
$U_0 \in C^2(\R^2)$, and the second derivatives of
$U_0$ are H\"older continuous.
\end{theorem}

\prf
As a first step we wish to show that $U_0$ and $\rho_0$ are bounded. Recall that
\[
-U_0(r)=4\int_0^\infty \frac{s}{r+s} \rho_0(s)\, K(\xi)\,ds
=I_1 + I_2,\ r \geq 0,
\]
where in $I_1$ the variable $s$ ranges in $[0,2r]$ and in 
$I_2$ it ranges in $[2r,\infty[$.
Using the estimate (\ref{kest}) and the fact that $\rho_0 \in L^{3/2}(\R^2)$
we find
\beas
I_1 
&\leq& 
\frac{C}{r} \int_0^{2r} s \rho_0 (s) \Bigl(1-\ln(1-[r,s])\Bigr)\,ds\\
&\leq&
\frac{C}{r} \nn{\rho_0}_{3/2} \left(\int_0^2r s
\Bigl(1-\ln(1-[r,s])\Bigr)^3 ds\right)^{1/3} \leq C r^{-1/3} .
\eeas
For $s \geq 2 r$ the elliptic integral $K(\xi)$ is bounded, and again
by H\"older's inequality we immediately obtain the same estimate for $I_2$ so 
that
\[
\n{U_0(r)} \leq C r^{-1/3},\ r>0.
\]
Next we know that
\[
\rho_0 (r) = \int q \left(E_0 - \frac{1}{2} v^2 - U_0(r) \right)\, dv
= \left\{ 
\begin{array}{ccl}
\displaystyle 2 \pi \int_{U(r)}^{E_0} q(E_0-E)\,dE &,& U_0(r) < E_0,\\
0 &,& U_0(r)\geq E_0 .
\end{array}
\right.
\]
The additional assumption on $Q'$ implies that there are constants
$C>0,\ \epsilon_0 >0$ such that
\[
q(\epsilon) \leq C \epsilon^{\mu_1},\ \epsilon \geq \epsilon_0,
\]
and this implies that 
\[
\rho_0 (r) \leq C r^{-(\mu_1+1)/3},\ r>0 .
\]
Since we know that $\rho_0$ has compact support, it follows that 
$\rho_0 \in L^3 (\R^2)$. We may now repeat the estimate for $U_0$
and obtain $I_1 \leq C r^{1/3}$ and $I_2 \leq C$ so that $U_0$ and
thus also $\rho_0$ are bounded. 

For the rest of our argument we rely on the regularity properties
of potentials generated by single layers. Firstly, the boundedness
of $\rho_0$ implies that $U_0$ is H\"older continuous, cf.\
\cite[page 42]{gue}. The relation between $\rho_0$ and $U_0$
immediately implies that $\rho_0$ shares this property. This
implies that $U_0$ has  H\"older continuous first order derivatives,
a fact known as Ljapunov's Theorem,
cf.\ \cite[pages 66,\ 67]{gue}. Since
\[
\rho_0'(r) = - 2 \pi q (E_0 -U_0(r))\, U_0'(r)
\]
$\rho_0$ is continuously differentiable. 

If $q$ is locally H\"older
continuous, then $\rho_0$ will have H\"older continuous
first order derivatives; note that $E_0-U_0(r)$ ranges in a bounded
interval for $r\in [0,\infty[$ so the local H\"older continuity
of $q$ suffices. We can now apply Ljapunov's Theorem again and
obtain the remaining assertions. \prfe

We remark that above we considered $U_0$ as a function on the
$(x_1,x_2)$ plane. Of course the definition (\ref{potential})
makes perfect sense on all of $\R^3$, and as long as we consider
only derivatives parallel to the $(x_1,x_2)$ plane all the regularity
assertions for $U_0$ hold on the whole space $\R^3$.
However, it is well known that the derivative of $U_0$
perpendicular to the plane has a jump discontinuity on the plane.

\end{document}